\documentclass[11pt]{article}
\usepackage{sidecap}
\usepackage{moriond,epsfig}
\usepackage{amssymb}
\usepackage[]{amsmath}

\def\be{\begin{equation}}
\def\ee{\end{equation}}
\def\bea{\begin{eqnarray}}
\def\eea{\end{eqnarray}}

\begin{document}
\vspace*{4cm}
\title{Supersymmetric Lepton Flavor Violation
and Leptogenesis}

\author{S.~Albino$^{\rm a}$, F.~Deppisch$^{\rm b}$, R.~R\"uckl$^{\rm a}$}

\address{$^{\rm a}$Institut f\"ur Theoretische Physik und Astrophysik,
Universit\"at W\"urzburg, D-97074 W\"urzburg, Germany\\
$^{\rm b}$Deutsches Elektronen-Synchrotron DESY, D-22603 Hamburg,
Germany}

\maketitle\abstracts{
We present and discuss constraints on supersymmetric
type I seesaw models imposed by neutrino data, charged lepton flavor
violation and thermal leptogenesis.
}

%%%%%%%%%%%%%%%%%%%%%%%%%%%%%%%%%%%%%%%%%%%%%%%%%%%%%%%%%%%%%%%%%%%%%
\section{Supersymmetric seesaw mechanism and slepton mass matrix}

The observed neutrino oscillations imply the existence of neutrino 
masses and lepton flavor mixing, and give hints towards physics 
beyond the Standard Model. For example, the smallness of the 
neutrino masses suggests the realization of the seesaw mechanism
involving heavy right-handed Majorana neutrinos. The latter  
violate lepton number and $CP$, and thus allow for leptogenesis.
Particularly interesting are supersymmetric scenarios, 
where the lepton flavor violation (LFV) present in the neutrino 
sector is transmitted to the slepton sector giving also rise to 
measurable LFV processes of charged leptons.

A minimal model of this kind is obtained if three right-handed neutrino singlet 
fields $\nu_R$ are added to the MSSM particle content. In this model, 
one can have the following Majorana mass and Yukawa interaction terms
\cite{Casas:2001sr}:
\begin{equation}
-\frac{1}{2}\nu_R^{cT} M \nu_R^c + \nu_R^{cT} Y_\nu L \cdot H_2, 
\label{suppot4}
\end{equation}
where $M$ is the Majorana mass matrix, \(Y_\nu\) is the matrix of Yukawa 
couplings, and $L$ and $H_2$ denote the left-handed lepton and hypercharge 
+1/2 Higgs doublets, respectively. Electroweak symmetry breaking then 
generates the neutrino Dirac mass matrix \(m_D=Y_\nu \langle H_2^0 \rangle\) 
where \(\langle H_2^0 \rangle = v\sin\beta\) is the appropriate Higgs v.e.v.\ 
with \(v=174\)~GeV and 
\(\tan\beta =\frac{\langle H_2^0\rangle}{\langle H_1^0\rangle}\).
If the mass scale $M_R$ of the matrix $M$ is much greater than the 
electroweak scale, and thus much greater than the scale of \(m_D\), 
one naturally obtains three light neutrinos with the mass matrix 
\begin{equation}\label{eqn:SeeSawFormula}
M_\nu = m_D^T M^{-1} m_D = Y_\nu^T M^{-1} Y_\nu (v \sin\beta )^2,
\end{equation}
and three heavy neutrinos with the mass matrix $M_N = M$.
In the basis assumed, M is diagonal, while $M_{\nu}$ is to be
diagonalized by the unitary MNS matrix \(U\):
\begin{eqnarray}\label{eqn:NeutrinoDiag}
U^T M_\nu U \!\!\! &=& \!\!\! \textrm{diag}(m_1, m_2, m_3),\\
U \!\!\! &=& \!\!\! V_\text{CKM}(\theta_{12}, \theta_{13}, \theta_{23},
\delta)\cdot \textrm{diag}(e^{i\phi_1},e^{i\phi_2},1), \nonumber
\end{eqnarray}
$\theta_{ij}$ being mixing angles, $\delta$ and $\phi_i\) being Dirac 
and Majorana phases, respectively, and $m_i$ being the
light neutrino mass eigenvalues.

The heavy neutrino mass eigenstates $N$, which are too heavy to be observed
directly, influence the evolution of the MSSM slepton mass matrix: 
\begin{eqnarray}
 m_{\tilde l}^2=\left(
    \begin{array}{cc}
        m_L^2    & m_{LR}^{2\dagger} \\
        m_{LR}^2 & m_R^2
    \end{array}
       \right)_{\rm MSSM}+\left(
    \begin{array}{cc}
       \delta m_L^2    & \delta m_{LR}^{2^\dagger} \\
        \delta m_{LR}^2 & \delta m_R^2
    \end{array}
      \right)_{\rm N}.
\end{eqnarray}
It is these flavor off-diagonal virtual effects which lead to 
charged LFV. Adopting the minimal supergravity (mSUGRA) scheme
one finds, in leading logarithmic approximation \cite{Hisano:1999fj},
\begin{equation}
  \delta m_{L}^2 = -\frac{1}{8 \pi^2}(3m_0^2+A_0^2)Y_\nu^\dag L Y_\nu,\;\;\;
  \delta m_{R}^2 = 0,\;\;\; 
  \delta m_{LR}^2 = -\frac{3}{16\pi^2} A_0 v \cos\beta Y_l Y_\nu^\dag L Y_\nu,
\label{left_handed_SSB2}
\end{equation}
where $L_{ij} = \ln(M_\text{GUT}/M_i)\delta_{ij}$, $M_i$ being
the heavy neutrino masses, and
\(m_0\) and \(A_0\) are the universal scalar mass and trilinear
coupling, respectively, at the GUT scale $M_\text{GUT}$. 

By inverting (\ref{eqn:SeeSawFormula}), the neutrino Yukawa matrix 
can be written as follows \cite{Casas:2001sr}:
\begin{eqnarray}\label{eqn:yy}
    Y_\nu \!=\!
    \frac{1}{v\sin\beta}\text{diag}(\sqrt{M_1},\sqrt{M_2},\sqrt{M_3})
    \!\cdot\!R\!\cdot\!\text{diag}(\sqrt{m_1},\sqrt{m_2},\sqrt{m_3})
    \!\cdot\! U^\dagger.
\end{eqnarray}
Here, a new complex orthogonal matrix $R$ appears which may be 
parametrized in terms of 3 complex angles $\theta_i=x_i +i y_i$:
\begin{equation}
R=\left(\begin{array}{ccc}
c_2 c_3 & -c_1 s_3-s_1 s_2 c_3 & s_1 s_3 -c_1 s_2 c_3\\
c_2 s_3 & c_1 c_3 -s_1 s_2 s_3 & -s_1 c_3 -c_1 s_2 s_3 \\
s_2 & s_1 c_2 & c_1 c_2 \end{array}\right),
\end{equation}
with $(c_i, s_i)=(\cos \theta_i,\sin \theta_i) =(\cos x_i \cosh
y_i -i \sin x_i \sinh y_i, \sin x_i \cosh y_i +i \cos x_i \sinh
y_i)$. 
While the light neutrino masses $m_i$ and the mixing angles 
$\theta_{ij}$ have been measured or at least constrained, the 
phases $\phi_i$ and $\delta$, the heavy neutrino masses $M_i$ 
and the matrix $R$ are presently unknown.  
Using the available neutrino data 
\cite{Maltoni:2003sr}
as input in the appropriate 
renormalization group equations, $Y_\nu$ is evolved from the 
electroweak scale to the GUT scale and then put into the 
renormalization of the slepton mass matrix from $M_{GUT}$
to the electroweak scale.

%%%%%%%%%%%%%%%%%%%%%%%%%%%%%%%%%%%%%%%%%%%%%%%%%%%%%%%%%%%%%%%%%%%%%%%
\section{Charged lepton flavor violation}

The renormalization effects (\ref{left_handed_SSB2}) lead to
charged LFV either via contributions of virtual sleptons in loops
such as in radiative decays \(l_i\to l_j\gamma\), or
via real slepton production and decay such as in 
$e^+e^- \to \tilde{l}_a^-\tilde{l}^+_b\to l_i^-l^+_j +
2\tilde{\chi}^0_1$. 
To lowest order in LFV couplings one has 
\cite{Casas:2001sr,Hisano:1999fj}
\begin{equation}\label{eqn:DecayApproximation}
\Gamma(l_i \to l_j \gamma) \propto \alpha^3 m_{l_i}^5
\frac{|(\delta m_L)^2_{ij}|^2}{\tilde{m}^8} \tan^2 \beta,
\end{equation}
$\tilde m$ characterizing the typical sparticle masses in the
loop. Similarly, one finds \cite{Deppisch:2003wt}
\begin{equation}\label{eqn:HighEnergyApproximation}
    \sigma(e^+e^- \to l_i^- l_j^+ +2\tilde\chi^0_1)
    \approx
        \frac{|(\delta m_L)^2_{ij}|^2}{m^2_{\tilde l}
        \Gamma^2_{\tilde l}}
        \sigma(e^+e^- \rightarrow l_i^- l_i^+
        +2\tilde\chi^0_1).
\end{equation}
Consequently, light neutrino data imply interesting constraints on LFV
processes through the dependence on $|(Y_\nu^\dagger L Y_\nu)_{ij}|^2$ 
with $Y_\nu$ given by Eq.\ (\ref{eqn:yy}).  
Conversely, measurements or bounds on LFV processes can constrain
the fundamental seesaw parameters $M_i$ and $R$. 

\begin{figure}[h!]
\centering
\begin{minipage}{.49\linewidth}\includegraphics[clip,width=0.9\textwidth]{low_SPS1.eps} 
\end{minipage} \hfill
\begin{minipage}{.49\linewidth}
\includegraphics[clip,width=0.92\textwidth]{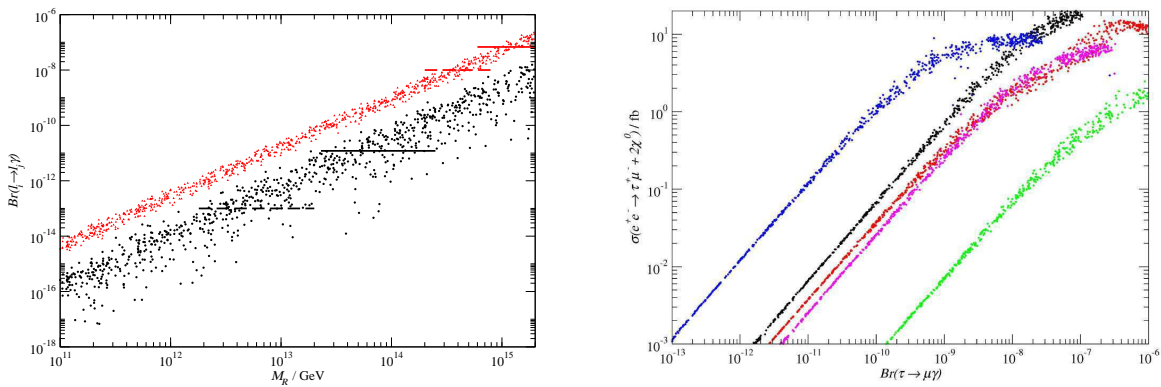}
\end{minipage}
\caption{(a) Br$(\tau \rightarrow \mu\gamma)$ (upper points) and 
Br$(\mu \rightarrow e\gamma)$ (lower) versus $M_R$ in mSUGRA
scenario SPS1a for real $R$. 
The light neutrino masses are assumed quasi-degenerate. 
The mixing angles and the Dirac phase are scattered within the full ranges 
consistent with experiment.
The solid (dashed) horizontal lines mark the present
~\protect\cite{BABAR:2004wt,Eidelman:2004wt}
(expected future) bounds,
Br$(\tau \rightarrow \mu
\gamma)<6.8 \times 10^{-8}$ ($10^{-8}$) and Br$(\mu \rightarrow e
\gamma)<1.2 \times 10^{-11}$ ($10^{-13}$).
(From~\protect\cite{Deppisch:2002vz}.)
(b) $\sigma(e^+e^- \to \tau^+\mu^- + 2\tilde\chi^0_1)$
versus Br$(\tau \rightarrow \mu\gamma)$. The light neutrino parameters 
are scattered as in (a). The plots (from left to right) are calculated
in the mSUGRA scenarios C', B', SPS1a, G' and I'. 
(From~\protect\cite{Deppisch:2003wt}.)}
\label{br12vsMRdegRSPS1a}
\end{figure}

For simplicity, we first consider the case of mass degenerate heavy 
Majorana neutrinos with $M_i=M_R$. If $R$ is real, i.e. $y_i=0$ , then 
it will drop out from the product $Y_\nu^\dagger Y_\nu$ in this case as
do the Majorana phases \(\phi_1\) and \(\phi_2\), leaving $M_R$ and
$\delta$ as the only unconstrained parameters.
Fig.~\ref{br12vsMRdegRSPS1a}a shows the typical rise of 
Br$(l_i \to l_j \gamma)$ with $M_R^2$ suggested by 
Eq.\ (\ref{eqn:DecayApproximation}) for fixed light neutrino masses.
Also shown is the impact of the uncertainties in the neutrino data. 
From the present bound \cite{Eidelman:2004wt}
Br$(\mu \to e \gamma) < 1.2 \cdot 10^{-11}$ 
one can derive an upper limit on $M_R$ of order $10^{14}$ GeV.
Furthermore, from  Eqs.\ (\ref{eqn:DecayApproximation}) and 
(\ref{eqn:HighEnergyApproximation}) the uncertainties 
in the neutrino parameters are expected to 
drop out of the correlation of radiative decays
and scattering processes in the same LFV channel. This is demonstrated
in Fig.~\ref{br12vsMRdegRSPS1a}b for the $\tau\mu$ channel.
As can be seen, combined measurements of both processes provide decisive 
tests of the considered scenarios. 

\begin{figure}[h!]
\centering
\includegraphics[clip,width=0.50\textwidth]{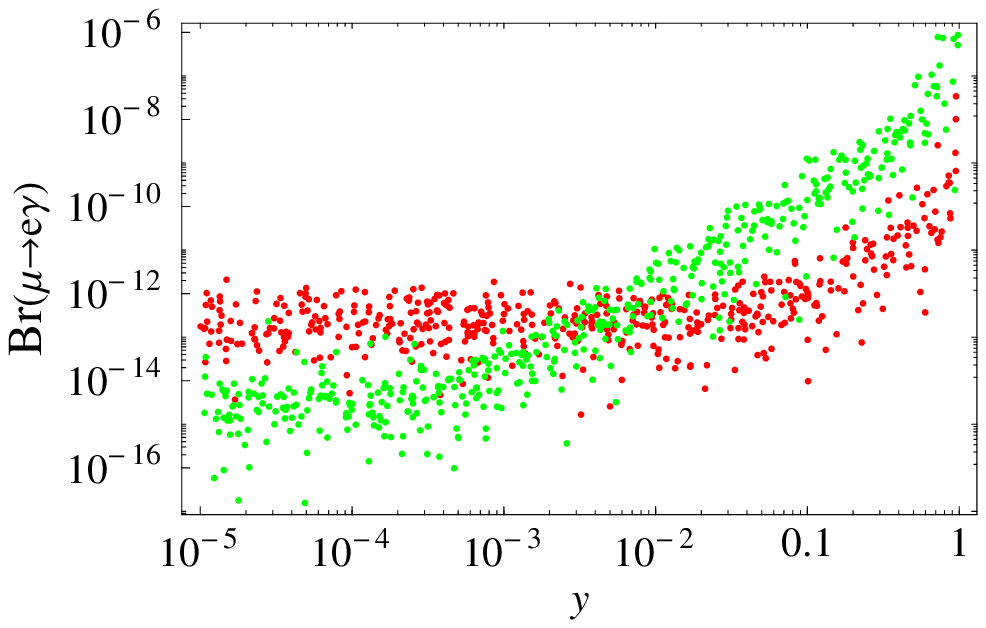}
\includegraphics[clip,width=0.49\textwidth]{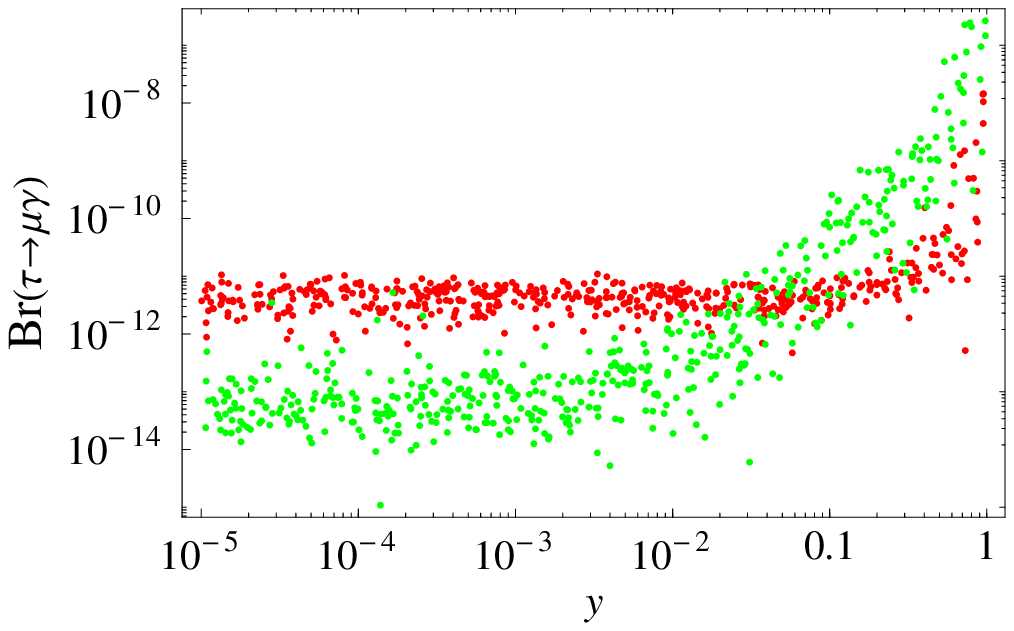}
\caption{Br$(l_i \rightarrow l_j \gamma)$ versus $y_i=y$ for fixed
$M_R =10^{12}$ GeV in mSUGRA scenario SPS1a for hierarchical
(dark points) and degenerate (light) light neutrino masses. The 
parameters are scattered as in Fig.~\ref{br12vsMRdegRSPS1a}a. In addition,
the $x_i$ are scattered over their full range $0<x_i<2 \pi$.}
\label{brijvsydegRfixMRscatallSPS1a}
\end{figure}

The above results are rather conservative, since LFV processes will be 
enhanced if the light neutrino masses are hierarchical 
instead of degenerate, and/or if $R$ is complex. In the latter case, 
LFV observables have rather more
freedom since the dependence on the $y_i$ can be as significant as
the $M_R$ dependence, as Fig.~\ref{brijvsydegRfixMRscatallSPS1a} shows.
The change in $Y_\nu^\dagger Y_\nu$ is approximately
\begin{equation}
\Delta_R (Y_\nu^\dagger Y_\nu)\approx U {\rm diag}(\sqrt{m_i})
(R^\dagger R-{\mathbf 1}){\rm diag}(\sqrt{m_i})U^\dagger.
\label{changeinYYfromy}
\end{equation}
Eq.\ (\ref{changeinYYfromy}) implies two features seen in 
Fig.~\ref{brijvsydegRfixMRscatallSPS1a}:
(i) Compared to the case of degenerate light neutrino masses, the $y$ 
dependence in the hierarchical case is weaker since the condition 
$m_3\gg m_{1,2}$
implies that only $(R^\dagger R-{\mathbf 1})_{33}= O(y_i^2)$ contributes. 
(ii) Since Eq.\ (\ref{changeinYYfromy}) is approximately imaginary and
linear in the $y_i$, non-zero $y_i$ can only increase the observable so that
lower limits obtained for real $R$ remain unaffected. 
Even small values of $y$ can enhance a process by orders of magnitude
from the real $R$ result. 
%Quantitatively, for
%neutrino parameters at their central values and vanishing phases, 
%Br$(l_i \rightarrow l_j \gamma) \propto
%A_{ij}^2+y^2$, with $A_{12(23)}\approx 10^{-4}(10^{-3})$ for 
%degenerate light neutrinos masses
%and $A_{12(23)}\approx 0.2$ for hierarchical ones. 
%Therefore, 
Also,
in contrast to the real $R$ case,
the LFV branching ratios for degenerate neutrinos can now be larger 
than that for hierarchical neutrinos. 

%%%%%%%%%%%%%%%%%%%%%%%%%%%%%%%%%%%%%%%%%%%%%%%%%%%%%%%%%%%%%
\section{Leptogenesis}

In thermal leptogenesis the baryon asymmetry of the universe 
is generated from out-of-equilibrium decays of the heavy 
Majorana neutrinos (see e.g. \cite{Buchmuller:2004nz,DiBari:2004en}). 
Most models are based on the assumption of a hierarchical
mass spectrum $M_1 \ll M_2 \ll M_3$.
It is then natural to assume that light neutrino masses are also hierarchical.
In this case, the baryon to photon ratio is determined by four factors
which are briefly explained below:
\begin{equation}\label{etaBshort}
    \eta_{B}
    \equiv
        \frac{n_B - n_{\bar{B}}}{n_\gamma}
    \approx
        d~a_{\rm Sph}~\epsilon_{1}~\kappa_{f}.
\end{equation}

The $CP$ asymmetry $\epsilon_1$ generated in the decays of the
lightest of the heavy Majorana neutrinos $N_1$ is 
\cite{Plumacher:1998ex,Davidson:2002qv}
\begin{equation}\label{eqn:CPAsymmetry}
    \epsilon_1
    \equiv
        \frac{\Gamma\left(N_{1}\to h_2 + l\right)-
          \Gamma (N_{1}\to \bar h_2 + \bar{l})}
          {\Gamma\left(N_{1}\to h_2 + l\right)+
          \Gamma (N_{1}\to \bar h_2 + \bar{l})}
    \approx
          -\frac{3}{8\pi}\frac{M_1}{v^2\sin^2\beta}\frac{\sum_i
          m_i^2 \text{Im} \left(R_{1i}^2\right)}{\sum_{i}
          m_i\left|R_{1i}\right|^2}.
\end{equation}
This relation clearly shows that non-zero imaginary parts of the \(R\)
matrix elements are necessary to generate a $CP$ asymmetry.
The efficiency factor \(\kappa_f\) in (\ref{etaBshort}) takes into
account the washout of the initial \((B-L)\) asymmetry. A reliable
numerical fit for \(\kappa_f\) for hierarchical light neutrinos in
the strong washout regime can be found in
\cite{Buchmuller:2004nz}. The solar and atmospheric neutrino mass
fits favor a value \(\kappa_f={\cal O}(10^{-2})\).
The \((B-L)\) asymmetry is subsequently converted to a baryon
asymmetry by sphaleron processes. In the case of the MSSM one
obtains the conversion factor \(a_{\rm Sph}=\frac{8}{23}\).
Finally, one has to take into account the dilution of the asymmetry due
to standard photon production, described by the dilution factor 
$d \approx \frac{1}{78}$ in the MSSM.
Confronted with the observed baryon asymmetry \cite{Tegmark:2003ud}
$\eta_B=(6.3\pm 0.3)\cdot 10^{-10}$,
relation (\ref{eqn:CPAsymmetry})
implies a lower bound on the \(M_1\) scale \cite{Davidson:2002qv},
e.g. if \(\epsilon_1>10^{-6}\), then \(M_1> 4\cdot 10^{9}\)~GeV.
Furthermore,
to allow for thermal production of right-handed neutrinos after inflation,
we require $M_1<10^{11}$ GeV, the maximum order of magnitude
that the reheating temperature can reach without suffering an overabundance of 
gravitinos, whose decay into energetic photons can otherwise spoil big bang nucleosynthesis.
\begin{figure}[t]
\centering
\includegraphics[clip,width=0.49\textwidth]{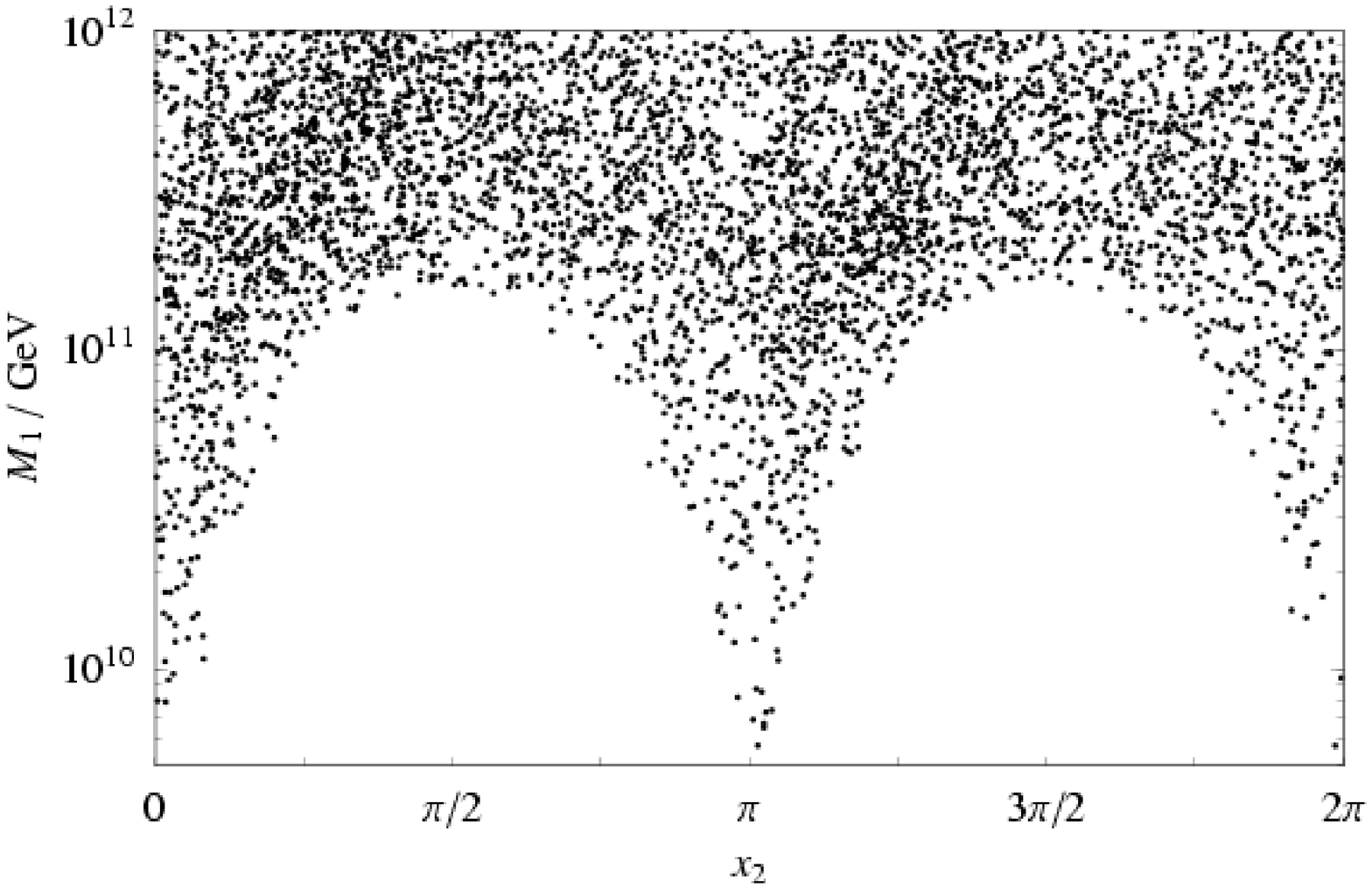}
\includegraphics[clip,width=0.49\textwidth]{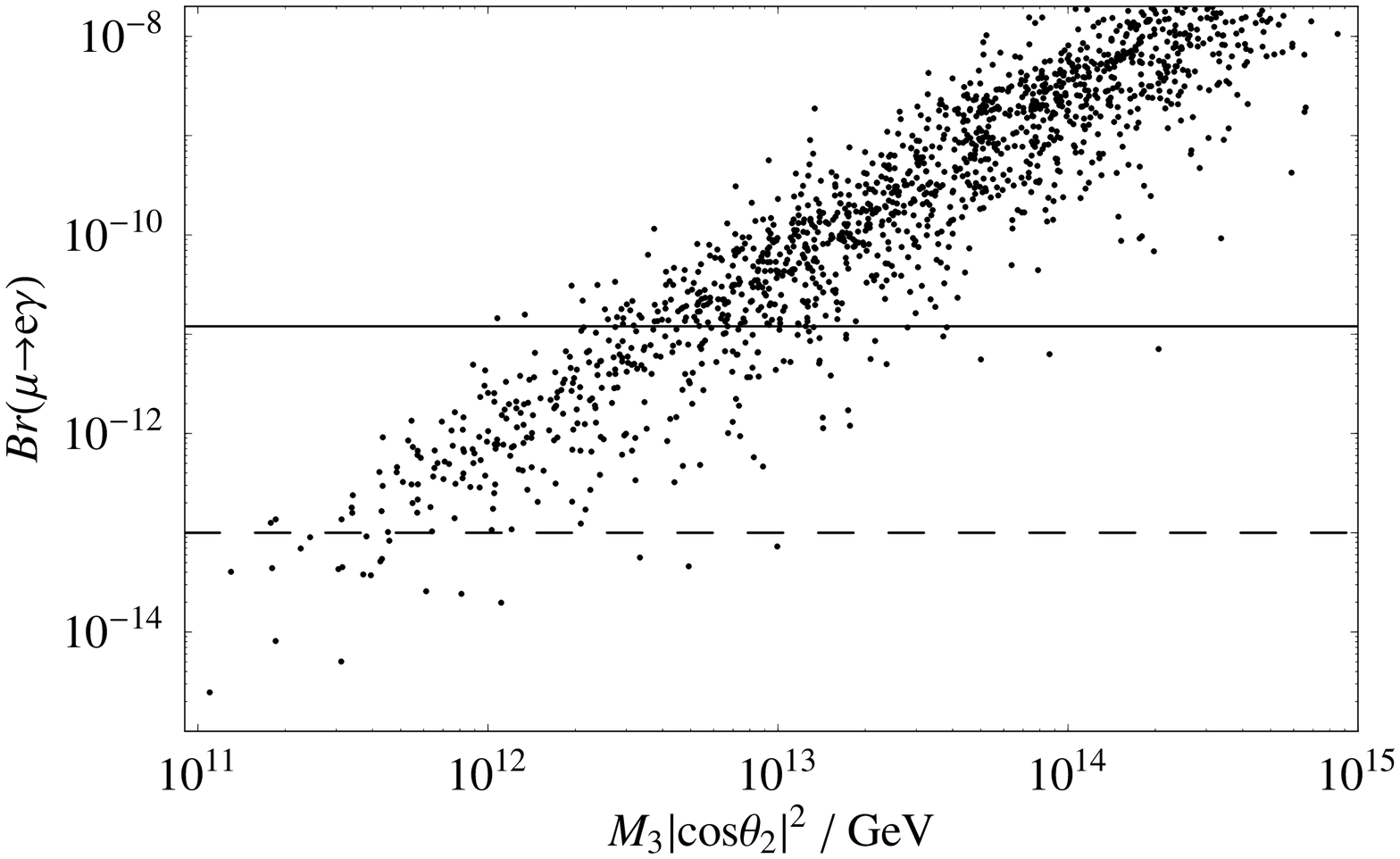}

\caption{(a) Region in the plane $(x_2, M_1)$ consistent with the
   generation of the baryon asymmetry \(\eta_B=(6.3\pm 0.3)\cdot 10^{-10}\)
   via leptogenesis. (b) \(Br(\mu\to e\gamma)\) as a function of
%            the heaviest Majorana mass 
             $M_3$ in mSUGRA scenario SPS1a, for
            \(M_1=10^{10}\)~GeV. The solid
            (dashed) line indicates the present (expected future) experimental
            sensitivity.
   All other seesaw parameters are scattered in their allowed
   ranges for hierarchical light and heavy neutrinos.}
     \label{fig:torboegen}
\end{figure}
In the above mass range,
the condition  to reproduce the experimental baryon asymmetry
then puts constraints on the parameters \(x_2\) and \(x_3\) of the \(R\)
matrix~\cite{Deppisch:2005rv} as illustrated in
Fig.~\ref{fig:torboegen}a. 
With decreasing $M_1<10^{11}$~GeV, the allowed
values for $x_2$ are concentrated more and more 
at $\sin x_2=0$. A similar 
behaviour is found for the angle $x_3$.

As discussed in the previous section for degenerate Majorana
neutrinos, experimental limits on \(Br(\mu\to e\gamma)\) can be used
to constrain the heavy neutrino scale, here represented by the
heaviest Majorana neutrino mass \(M_3\). This is shown in
Fig.~\ref{fig:torboegen}b. The present bound on $Br(\mu\to e
\gamma)$ constrains $M_3$ already to be smaller than 
$10^{13}$~GeV. Future experiments are expected to 
reach below $M_3 = 10^{12} $~GeV.

%%%%%%%%%%%%%%%%%%%%%%%%%%%%%%%%%%%%%%%%%%%%%%%%%%%%%%%%%%%%%
\section{Concluding remarks}
 
\begin{SCfigure}[][h!]
\centering
\includegraphics[height=6cm]{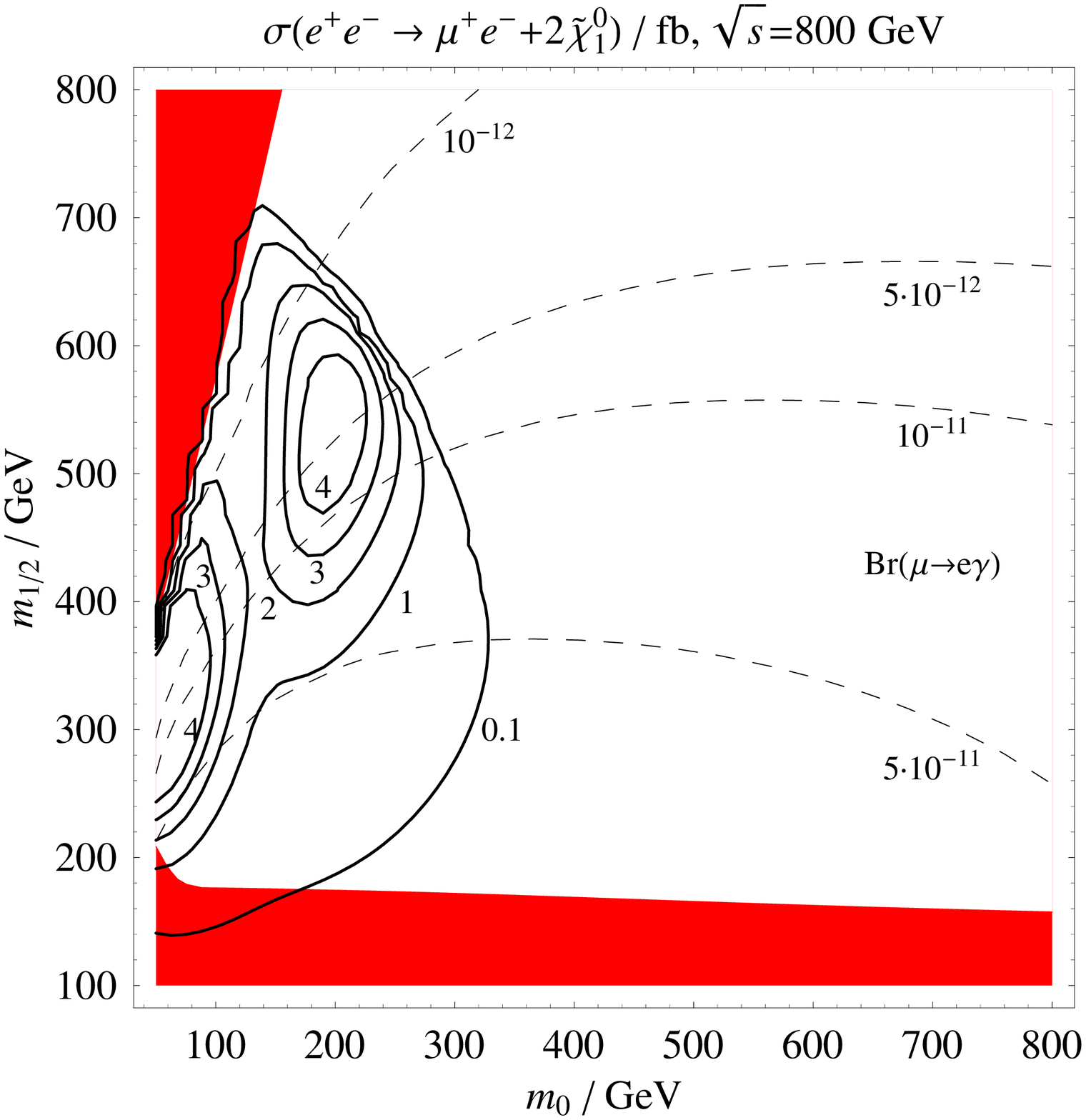}
\caption{Contours of the polarized cross section
\(\sigma(e^+e^-\to\mu^+e^- +2\tilde\chi_1^0)\) at
\(\sqrt{s}_{ee}=800\) GeV in the \(m_0-m_{1/2}\) plane (solid
lines). The remaining mSUGRA parameters are: \(A_0=0\)~GeV,
\(\tan\beta=5\), \(\text{sign}(\mu)=+\). The beam polarizations
are: $P_{e^-} = +0.9, P_{e^+} = +0.7$. For comparison, \({\rm
Br}(\mu\to e\gamma)\) is shown by dashed lines. The neutrino
oscillation parameters are fixed at their best fit values, the
lightest neutrino mass and all complex phases are set to
zero, and the degenerate Majorana mass scale is
\(M_R=10^{14}\)~GeV. The shaded (red) areas are forbidden by mass
bounds from various experimental sparticle searches.}
\label{fig:scan_ep1}
\end{SCfigure}

Supersymmetric LFV is very model dependent.
As far as the mSUGRA parameters are concerned, 
%For example,
%as illustrated in Fig.~\ref{fig:scan_ep1},
%the ratio of two observables which occur through the same
%flavor mixing channel, although approximately independent of the 
%seesaw parameters, still depend crucially on the mSUGRA parameters. 
this is further emphasized in Fig. ~\ref{fig:scan_ep1}, where 
a scan in the $(m_0,m_{1/2})$ plane is performed showing contours
of fixed LFV cross section at the ILC and $\mu \to e \gamma$ 
branching ratio. Clearly, the strategy to probe LFV 
considered here will only be applicable once sufficient 
measurements of the SUSY particles' properties have been made.

\begin{SCfigure}[][h!]
\centering
\includegraphics[bb=88 32 338 187,height=5cm]{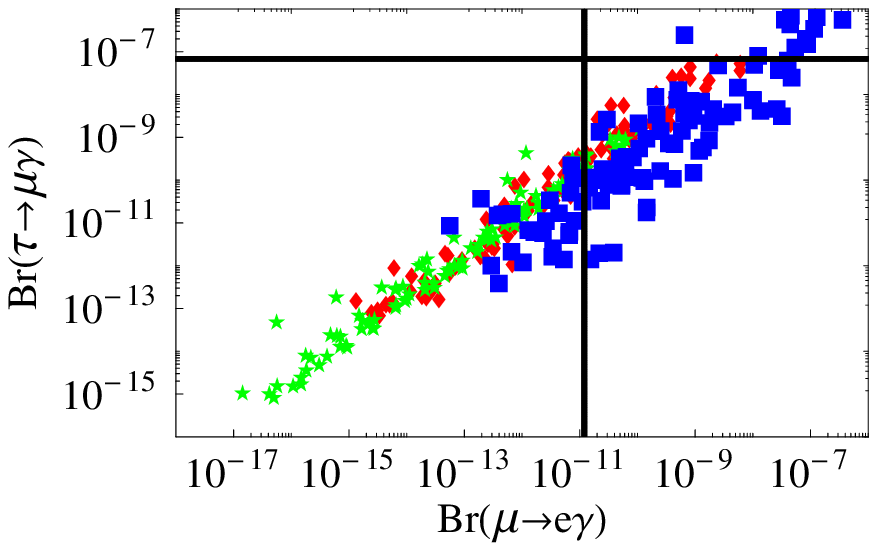}
\caption{Br$(\tau \rightarrow \mu \gamma)$ versus Br$(\mu
\rightarrow e\gamma)$ in mSUGRA scenario SPS1a, 
assuming degenerate Majorana masses
and hierarchical (diamonds) and degenerate (stars)
light neutrino masses, with 
parameters scattered as in Fig.~\ref{br12vsMRdegRSPS1a}.
In the case of hierarchical Majorana and light neutrino masses (squares)
all seesaw parameters including the
$x_i$ are scattered within their experimentally allowed ranges,
while the $y_i$ and $M_i$ are scattered within the bounds demanded
by leptogenesis and perturbativity.
Also indicated are the present experimental bounds
~\protect\cite{Eidelman:2004wt,BABAR:2004wt}.} 
\label{br12v23SPS1aallRallL2}
\end{SCfigure}

In a given scenario LFV observables, particularly in the 
$\mu e$ channel, are very sensitive to uncertainties in the 
neutrino parameters. However, correlations of observables in
the same LFV channel suffer much less from this.
As demonstrated in Fig.~\ref{br12vsMRdegRSPS1a}b, such correlations 
can be rather strong, and therefore very useful probes. 
Although less pronounced, relations also exist among observables 
in different LFV channels like $\tau\mu$ and $\mu e$.
These can be exploited to improve direct bounds. 
Fig.~\ref{br12v23SPS1aallRallL2} illustrates this for a variety of 
seesaw parameters, uncovering a rather good correlation between 
the $\mu e$ and $\tau \mu$ channels. One sees that  
the present experimental bound on Br$(\mu\rightarrow e\gamma)$
can be used to improve the direct bound on Br$(\tau \rightarrow \mu \gamma)$
by almost 2 orders of magnitude. Interestingly, 
this result does not depend on whether
hierarchical or degenerate heavy and light neutrinos are assumed. 

For hierarchical Majorana masses, leptogenesis provides
additional constraints. For a heavy Majorana neutrino mass spectrum 
obeying \(10^{10}~{\rm GeV} \approx M_1 \ll M_2 \ll M_3 \approx
10^{13}~{\rm GeV}\) (in mSUGRA scenario SPS1a), 
both LFV and leptogenesis are found to be viable.
The heaviest Majorana mass $M_3$ is constrained from above 
by the LFV process $\mu\to e\gamma$, the lightest mass $M_1$
from below by leptogenesis.
%For hierarchical Majorana masses, leptogenesis provides an
%additional constraint. The lightest right handed neutrino mass
%$M_1$ is fixed by the CP asymmetry $\epsilon_1$ imposed to
%generate the observed baryon asymmetry and the requirement of a
%small reheating temperature. The heaviest Majorana mass $M_3$ is
%constrained from above by the LFV process $\mu\to e\gamma$. Thus
%the right handed neutrino spectrum can be summarized as
%\(10^{11}~{\rm GeV} \approx M_1 \ll M_2 \ll M_3 \approx
%10^{13}~{\rm GeV}\), in the given mSUGRA scenario SPS1a.
%
In addition, the orthogonal $R$
matrix encoding the mixing of the right-handed neutrinos, which is
parametrized by the angles $\theta_i = x_i + i y_i$,
is constrained according to
\(\sin x_{2,3} \simeq 0\) and \(y_i <{\cal O}(1)\) as a consequence of
the successful leptogenesis condition for small
$M_1$ and the requirement of perturbative Yukawa couplings, respectively.
Finally, the remaining parameter $x_1$ can be constrained from 
the ratio $Br(\mu\to e\gamma)/Br(\tau\to\mu\gamma)$ 
\cite{Deppisch:2005rv}. 
Other work along this line can be found in the literature (for references 
see \cite{Deppisch:2005rv}) and has also been presented at this meeting
(see e.g. \cite{Petcov:2005rv}).

\section*{Acknowledgments}
We thank H. P\"as, A. Redelbach and Y. Shimizu for fruitful
collaboration. This work
was supported by the Federal Ministry of Education and Research
(BMBF) under contract number 05HT1WWA2.

\newpage
%
%In Figure~\ref{fig:scan_ep1} we show contour plots of the
%polarized cross section $\sigma(e^+e^-\to\mu^+ e^- +
%2\tilde\chi^0_1)$ in the $m_0-m_{1/2}$ plane. For comparison we
%overlay the branching ratio of $\mu\to e\gamma$ by dashed lines.
%Due to its radiative nature, \(\mu\to e\gamma\) generally has a
%better sensitivity for large sparticle masses (current limit:
%\(Br(\mu\to e\gamma)<10^{-11}\)), but the collider signal can be
%more sensitive in the large \(m_{1/2}\), small \(m_0\) region.
\vspace{-1cm}

\end{document}